\documentclass[aps,prd,preprint,superscriptaddress,tightenlines,nofootinbib,showpacs]{revtex4}

\usepackage{graphicx}
\usepackage{dcolumn}
\usepackage{bm}

\begin{document}

\preprint{CLNS 03/1826}       
\preprint{CLEO 03-09}         

\title{Observation of a Narrow Resonance of Mass 2.46 GeV/c$^2$ \\
       Decaying to \boldmath $D_s^{*+}\pi^0$  
       and Confirmation of the $D_{sJ}^{*}(2317)$ State}

%
\thanks{Submitted to Physical Review D; supersedes 
        hep-ex/0305017.}
%
%
\author{D.~Besson}
\affiliation{University of Kansas, Lawrence, Kansas 66045}
\author{S.~Anderson}
\author{V.~V.~Frolov}
\author{D.~T.~Gong}
\author{Y.~Kubota}
\author{S.~Z.~Li}
\author{R.~Poling}
\author{A.~Smith}
\author{C.~J.~Stepaniak}
\author{J.~Urheim}
\affiliation{University of Minnesota, Minneapolis, Minnesota 55455}
% ...

\collaboration{CLEO Collaboration} 
\noaffiliation


\date{May 28, 2003}

\begin{abstract} 
  Using 13.5 fb$^{-1}$ of $e^+e^-$ annihilation data collected with 
the CLEO II detector, we have observed a narrow resonance decaying  
to $D_s^{*+}\pi^0$, with a mass near 2.46 GeV/c$^2$. 
The search for such a state was motivated by the recent discovery 
by the BaBar Collaboration of a narrow state at 
2.32 GeV/c$^2$, the $D_{sJ}^*(2317)^+$, that decays to $D_s^+\pi^0$.  
Reconstructing the $D_s^+\pi^0$ and $D_s^{*+}\pi^0$ final states 
in CLEO data, we observe peaks in both of the corresponding 
reconstructed mass difference distributions, 
$\Delta M(D_s\pi^0) = M(D_s\pi^0) - M(D_s)$ and 
$\Delta M(D_s^*\pi^0) = M(D_s^*\pi^0) - M(D_s^*)$, 
both of them at values near 350 MeV/c$^2$.  We interpret these peaks 
as signatures of two distinct states, the $D_{sJ}^*(2317)^+$ 
plus a new state, designated as the $D_{sJ}(2463)^+$. 
Because of the similar $\Delta M$ values, each of these states 
represents a source of background for the other if photons are 
lost, ignored or added.  A quantitative accounting of these reflections 
confirms that both states exist.
We have measured the mean mass differences 
$\langle{\Delta M}(D_s\pi^0)\rangle = 
350.0\pm 1.2\;\mbox{\rm [stat.]}\pm 1.0\;\mbox{\rm [syst.]}\;$MeV/c$^2$ 
for the $D_{sJ}^*(2317)^+$ state, and 
$\langle{\Delta M}(D_s^*\pi^0)\rangle = 
351.2\pm 1.7\;\mbox{\rm [stat.]}\pm 1.0\;\mbox{\rm [syst.]}\;$MeV/c$^2$ 
for the new $D_{sJ}(2463)^+$ state. 
We have also searched, but find no evidence, for decays of the 
two states via the channels  
$D_s^{*+}\gamma$, $D_s^+\gamma$, and $D_s^+\pi^+\pi^-$.
The observations of the two states at 2.32 and 2.46 
GeV/c$^2$, in the $D_s^+\pi^0$ and $D_s^{*+}\pi^0$ decay channels 
respectively, are consistent with their 
interpretations as $c\overline{s}$ mesons with orbital angular 
momentum $L=1$, and spin-parity $J^P = 0^+$ and $1^+$.
\end{abstract}

\maketitle


\section{Introduction}

  The BaBar Collaboration has recently reported~\cite{babar} 
evidence for a new narrow resonance with a mass near 2.32\ GeV/c$^2$, 
which decays to $D_s^+ \pi^0$.  The BaBar data are 
consistent with the identification of this state as one of the four 
lowest-lying 
members of the $c\overline{s}$ system with orbital angular momentum 
$L=1$, and provisionally it has been named the $D_{sJ}^*(2317)$ meson.  
A natural candidate would be the $^3P_0$ $c\overline{s}$ meson 
with spin-parity $J^P = 0^+$, but other possibilities, 
including exotic states, are not ruled out.  
In this paper, we report on a search for the $D_{sJ}^*(2317)$ meson,  
as well as other, possibly related states, 
in data collected with the CLEO II detector in symmetric 
$e^+e^-$ collisions at the Cornell Electron Storage Ring, at 
center-of-mass energies $\sqrt{s} \approx 10.6$\ GeV.  

  The spectroscopy of $P$-wave $c\overline{s}$ mesons is 
summarized in Ref.\ \cite{bartelt}.  Prior to the observation of the 
$D_{sJ}^*(2317)$, theoretical expectations~\cite{rgg,gi,iw,gk,dipe} 
were that: 
(1) all four states with $L=1$ are massive enough that their dominant 
    strong decays would be to the isospin-conserving $DK$ and/or $D^*K$ 
    final states, 
(2) the singlet and triplet $J^P=1^+$ states could mix, and 
(3) in the heavy quark limit, the two states with $j=3/2$ would be 
    narrow while the two with $j=1/2$ would be broad, where $j$ is 
    the sum of the strange quark spin and the orbital angular momentum.
Existing experimental evidence~\cite{pdg,cleods2} for the narrow 
$D_{s1}(2536)$ and $D_{sJ}^*(2573)$ mesons which decay dominantly 
to $D^*K$ and $DK$ respectively, and the compatibility 
of the $D_{sJ}^*(2573)$ with the $J^P$ assignment as $2^+$ support 
this picture.  

  The observation by BaBar \cite{babar} of the new state at a 
mass of 2.32 GeV is surprising because: 
(1) it is narrow (with intrinsic width $\Gamma < 10\;$MeV), 
(2) it has been observed in the isospin-violating $D_s \pi^0$ channel, 
and (3) its mass ($2316.8\pm 0.4\;\mbox{\rm [stat.]}\;$MeV/c$^2$) is 
smaller than most theoretical predictions for a $0^+$ $c\overline{s}$ 
state that could decay via this channel.  However, points 
(1) and (2) would be obvious consequences of the low mass, since the 
$D^{(*)}K$ decay modes are not allowed kinematically.  We 
also note that at least two theoretical calculations~\cite{nrz,bh} 
prior to the $D_{sJ}^*(2317)^+$ observation had suggested 
that, in the heavy quark limit, the $j=1/2$ states with 
$J^P = 0^+$ and $1^+$ 
could be thought of as chiral partners of the $D_s$ and $D_s^*$ 
mesons, and thus would be relatively light.  In one model~\cite{bh}
it was proposed that the mass splittings between the $0^+$ and $0^-$ 
states of heavy flavored mesons could be as small as 338 MeV/c$^2$, 
which is near the $D_{sJ}^*(2317)^+ - D_s^+$ mass 
splitting of $348.3\;$MeV/c$^2$ measured by BaBar.

Since the initial observation, a number of explanations have 
appeared~\cite{cahnjackson,barnescloselipkin,beverenrupp,
bardeeneichtenhill,szczepaniak,godfrey,colangelodefazio,bali}. 
Cahn and Jackson~\cite{cahnjackson} 
apply non-relativistic vector and scalar exchange 
forces to the constituent quarks. Barnes, Close and 
Lipkin~\cite{barnescloselipkin} consider 
a quark model explanation unlikely and propose a $DK$ molecular state. 
Similarly, Szczepaniak~\cite{szczepaniak} suggests a $D\pi$ atom.
Also going beyond a simple quark model description, 
Van Beveren and Rupp~\cite{beverenrupp} present 
arguments for a low mass $0^+$ $c\overline{s}$ state 
based on a unitarized meson model, 
by analogy with members of the light scalar meson nonet. 
Bali~\cite{bali} reports on lattice QCD calculations that predict 
signficantly larger $0^+ - 0^-$ meson mass splittings than what has 
been observed for the $D_{sJ}^*(2317) - D_s$ splitting.

On the contrary, Bardeen, Eichten and Hill~\cite{bardeeneichtenhill}  
couple chiral perturbation theory with a quark model 
representation in heavy quark effective theory, 
building on the model described in Ref.~\cite{bh}. 
They infer that the $D_{sJ}^*(2317)$ is indeed the $0^+$ $c\overline{s}$ 
state expected in the quark model, predict the existence 
of the $1^+$ partner of this state with a $1^+ - 1^-$ mass 
splitting equal to the $0^+ - 0^-$ mass splitting, and 
compute the partial widths for decays to allowed final states.
Godfrey~\cite{godfrey} and Colangelo and De~Fazio~\cite{colangelodefazio}
find that the radiative transistion of the $D_{sJ}^*(2317)$ should be 
significant if it is indeed a $c\overline{s}$ state.

  The goals of the analysis presented here are to use CLEO data to 
provide independent evidence regarding the existence of 
the $D_{sJ}^*(2317)$, to shed additional light on its properties, 
and to search for decays of other new, possibly related states.    
In particular, we address the following questions.  
Are the electromagnetic decays $D_s\gamma$ or $D_s^*\gamma$ 
observable in light of the isospin suppression of the strong decay 
to $D_s\pi^0\,$?  Are other strong decays observable such as 
$D_s^*\pi^0$, or the isospin-conserving but 
Okubo-Zweig-Iizuka (OZI) suppressed~\cite{ozi} 
decay $D_s\pi^+\pi^-\,$?  If the $D_{sJ}^*(2317)$ is the 
expected $0^+$ $c\overline{s}$ state, might the remaining $1^+$ state 
also be below threshold for decay to $D^*K$, as suggested in 
Ref.~\cite{bardeeneichtenhill},  
and thus be narrow enough to be observable in its decays to 
$D_s^*\pi^0$, $D_s\gamma$ or $D_s^{*}\gamma\,$? 

This article~\cite{cleohepex} is organized as follows.
After describing the detector and data set in Section~\ref{s-detector}, 
we summarize the reconstruction of the $D_{sJ}^*(2317)^+ \to D_s^+\pi^0$ 
decay channel in Section~\ref{s-conf}, including 
efforts to understand and exclude contributions from 
known background processes.  We then report in Section~\ref{s-searches} 
on searches for other possible decay channels as described 
in the preceding paragraph.  In Section~\ref{s-observe}, 
we report on the appearance of 
a statistically significant signal in the $D_s^{*+}\pi^0$ 
channel at a mass of 2.463 GeV/c$^2$, not compatible 
with a kinematic reflection of the $D_{sJ}^*(2317)^+$.  
We describe a quantitative analysis of the signals 
in the $D_s^{+}\pi^0$ and $D_s^{*+}\pi^0$ 
channels, leading us to infer the existence of two distinct 
states.  Based on this conclusion, we discuss the 
properties of these two states in Section~\ref{s-properties}, 
after which we summarize the principal results of the analysis.  

\section{Detector and Data Set}
\label{s-detector}

  The analysis described here is based on 13.5~fb$^{-1}$
of $e^+e^-$ collision data collected between 1990 and 1998.  
CLEO II is a general purpose, large solid angle, cylindrical 
detector featuring precision charged particle tracking and 
electromagnetic calorimetry, and is described in detail in 
Refs.\ \cite{Kubota:1992ww,Hill:1998ea}.  In its initial 
configuration, the tracking system was comprised of a six-layer 
straw tube chamber just outside of a 3.2 cm radius beryllium beam 
pipe, followed by a 10 layer hexagonal cell drift chamber and a 
51 layer square cell drift chamber, immersed in a $1.5\;$T magnetic 
field generated by a superconducting solenoid.  In 1995, the 
beam pipe and straw tubes were replaced by a 2.0 cm radius 
beam pipe plus three layers of silicon strip detectors each with 
double-sided readout, and a helium-propane gas mixture replaced 
the argon-ethane mixture previously used in the main drift chamber.  

  Beyond the tracking system, but within the solenoid, were also 
located a 5 cm thick plastic scintillation counter system for 
time-of-flight measurement and triggering, and a barrel calorimeter 
consisting of 6144 tapered CsI(Tl) crystals 30 cm in length, arrayed 
in a projective geometry, with their long axis oriented radially with 
respect to the $e^+e^-$ interaction point.  An additional 1656 
crystals were deployed in two end caps to complete the solid angle 
coverage.  The excellent energy and angular resolution of the 
calorimeter is critical for the reconstruction of 
$\pi^0\to\gamma\gamma$ decays as well as single low-energy photons 
such as those emitted in the $D_s^{*+}\to D_s \gamma$ transition.

\section{Confirmation of \boldmath $D^{*}_{sJ}(2317)^+ \to D_s^+ \pi^0$}
\label{s-conf}

  The search for the $D^{*}_{sJ}(2317)$ was carried out by 
reconstructing the $D_s^+ \pi^0$ state, using the $D_s^+ \to 
\phi\pi^+$ channel with $\phi\to K^+K^-$.  
Charge conjugation is implied throughout this article.
Pairs of oppositely charged 
tracks were considered as candidates for the decay products of 
the $\phi$ if the specific ionization ($dE/dx$) was measured 
in the main drift chamber to be within 2.5 standard deviations 
of the expectation for a kaon, and if the invariant mass of 
the $K^+K^-$ system was within $\pm 10\;$MeV/c$^2$ of the $\phi$ mass.
A third track with $dE/dx$ consistent with the expectation for a pion 
was combined with the $K^+K^-$ system to form a $D_s^+$ candidate 
with mass $M(KK\pi)$.
To improve resolution we adjust the momenta of the three particles 
subject to the constraint that their trajectories intersect at a 
common point corresponding to the decay point of a $D_s$ meson. 
When fitted to a Gaussian, the observed $D_s^+$ mass peak has a  
standard deviation ($\sigma$) 
of $6.5\pm 0.4\;$MeV/c$^2$ in our data, consistent with CLEO Monte Carlo 
simulations of $D_s$ production and decay plus a GEANT-3~\cite{geant} 
based simulation of particle propagation and detector response.

Clusters of energy deposition in the calorimeter unassociated 
with charged particle interactions were identified as potential 
photon candidates.  To be considered as candidates for the photons 
from $\pi^0\to \gamma\gamma$ decay, clusters with energy greater 
than 100 MeV located in the central region of the calorimeter 
($|\cos \theta| < 0.71$, where $\theta$ is measured with respect 
to the beam axis) were selected. 
Pairs of photons were required to satisfy 
$-3.0 < [M(\gamma\gamma) - M_{\pi^0}]/\sigma(\gamma\gamma) < 2.5 $, 
where $M(\gamma\gamma)$ is the invariant mass of the two photons
and $\sigma(\gamma\gamma)$ is the expected resolution on this mass.  
For each cluster being considered as a photon candidate, 
we additionally required that the lateral profile of energy deposition 
in the calorimeter be consistent, at the $99\%$ confidence level, 
with expectations for photons.  This requirement removes spurious 
photon candidates that are mainly due to inelastic interactions of 
charged hadrons or long-lived neutral hadrons.  The peak in the 
$M(\gamma\gamma)$ distribution for photon-pairs accompanying 
a $D_s^+$ candidate 
with $M(D_s) = M(KK\pi)$ between 1.9565 and 1.9805 GeV/c$^2$ 
has $\sigma = 5.8\pm 0.4\;$MeV/c$^2$ in our data, 
consistent with expectations from the Monte Carlo simulations.
Once identified as a $\pi^0$ candidate, the directions and  
energies of the two photons are adjusted with a kinematic fit 
to reconstruct to the known value~\cite{pdg} 
for the $\pi^0$ mass $M_{\pi^0}$.

\begin{figure}
  \includegraphics*[width=4.5in]{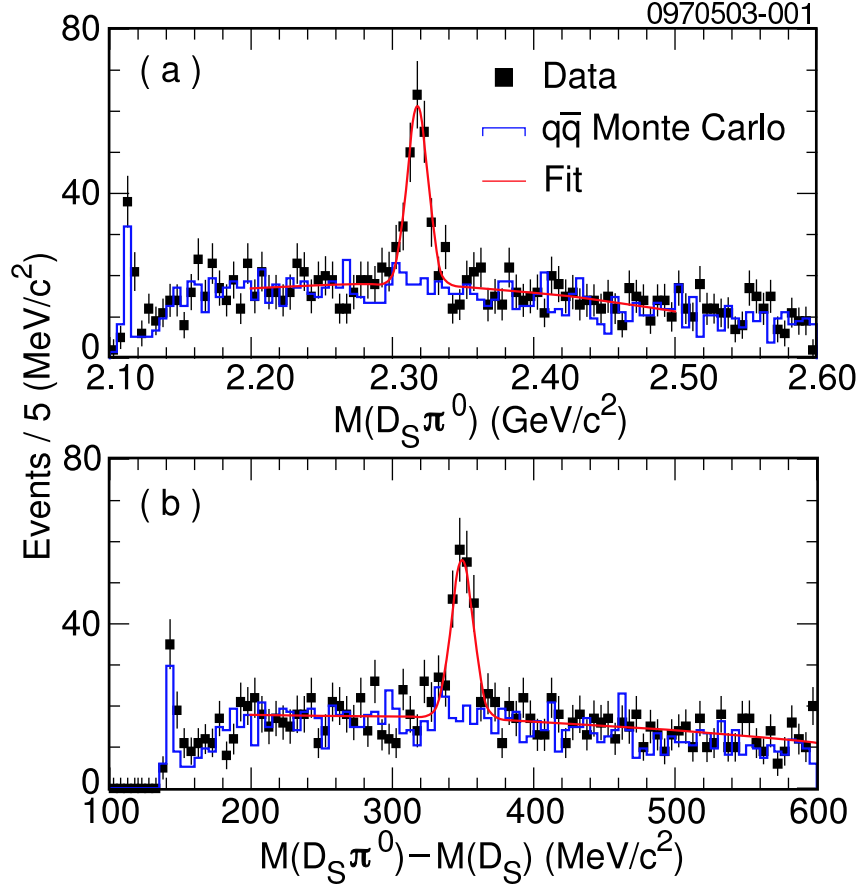}
  \caption{Distributions of (a) the masses $M(D_s\pi^0)$ 
           of the $D_s\pi^0$ candidates and (b) the 
           mass differences $\Delta M(D_s\pi^0) = M(D_s\pi^0) - M(D_s)$ 
           for events satisfying cuts on $M(KK\pi)$ consistent with the 
           $D_s$ mass
           and $M(\gamma\gamma)$ consistent with the $\pi^0$ mass, 
           as described in the text.  The points represent the CLEO data, 
           while the solid histogram is the predicted spectrum from the 
           Monte Carlo simulation of $e^+e^-\to q\overline{q}$ events.  
           The predicted spectrum is normalized absolutely  
           by the ratio of the equivalent luminosity of the Monte 
           Carlo sample used to the luminosity of the CLEO data sample.  
           The overlaid curve 
           represents the results from a fit of the data to a Gaussian 
           signal function plus a second-order polynomial background 
           function.
           }
  \label{fig:manddeltam}
\end{figure}

To suppress combinatoric backgrounds, we further required that the 
momentum of the $D_s^+\pi^0$ candidate be greater than $3.5\;$GeV/c.  
We also required that the helicity angle of the $\phi\to K^+K^-$ decay 
satisfy the requirement $|\cos{\theta_h}| > 0.3$, where $\theta_h$ is 
the angle between the $K^+$ momentum vector measured 
in the $\phi$ rest frame, and the $\phi$ momentum vector 
measured in the $D_s$ rest frame.  
The expected distribution from real $\phi$ decays varies as 
$\cos^2{\theta_h}$, whereas combinatoric backgrounds tend to be flat.
For $D_s\pi^0$ combinations satisfying these requirements, 
we plot the mass $M(D_s\pi^0) = M(KK\pi\pi^0)$ and the mass 
difference $\Delta M(D_s\pi^0) = M(D_s\pi^0) - M(D_s)$ 
in Fig.\ \ref{fig:manddeltam}(a) and (b), 
respectively. 
To improve the experimental resolution on $M(D_s\pi^0)$, 
the known value of the $D_s$ mass, 
$M_{D_s} = 1968.5\pm 0.6\;$MeV/c$^2$~\cite{pdg}, 
has been used to determine the 
energy of the $KK\pi$ system from its measured 
momentum in Fig.~\ref{fig:manddeltam}(a);
this substitution is not done for $\Delta M(D_s\pi^0)$ 
in Fig.~\ref{fig:manddeltam}(b), or for the 
calculation of other mass differences entering this 
analysis.

The narrow peaks in Fig.\ \ref{fig:manddeltam} at a mass 
near 2.32 GeV/c$^2$ and a $\Delta M(D_s\pi^0)$ near 350 MeV/c$^2$ are 
in qualitative 
agreement with the BaBar observation.  We note that there are no 
peaks in this region when $KK\pi$ combinations with $M(KK\pi)$ lying 
in $D_s$ side band regions are combined with a $\pi^0$.  The other feature 
of note in the spectra  
is the sharp signal from $D_s^{*+}\to D_s^+\pi^0$~\cite{cleodspi0}
near the kinematic threshold. 
In addition, Monte Carlo simulations of inclusive multi-hadron 
production via $e^+e^-\to q\overline{q}$ ($q=u,\,d,\,s,\,c$) 
give $M(D_s\pi^0)$ and $\Delta M(D_s\pi^0)$ 
spectra that reproduce the features observed in the data, except for 
the peaks near 2.32 GeV/c$^2$ and 350 MeV/c$^2$ in the respective plots.  
This is also illustrated in Fig.\ \ref{fig:manddeltam}, where the 
normalization for the $q\overline{q}$ Monte Carlo spectra is fixed 
by the ratio of the luminosity of the data sample to the equivalent 
luminosity of the Monte Carlo sample.  This normalization is known 
to a precision of approximately $\pm 5\%$.

The agreement between the Monte Carlo and data distributions 
in Fig.~\ref{fig:manddeltam} in normalization as well as shape
demonstrates that the simulation of `random' photons accompanying 
$D_s$ decays is accurate.  The accuracy of this simulation is  
important for our detailed analysis of this signal, 
described in Section~\ref{sec:feeddown}.

We have investigated mechanisms by which a peak at 2.32 GeV/c$^2$ 
could be generated from decays involving known particles, 
either through the addition, omission or substitution 
of a pion or photon, or through the mis-assignment of 
particle masses to the observed charged particles.  In no 
cases were narrow enhancements in the $M(D_s\pi^0)$ spectrum 
near 2.32 GeV/c$^2$ observed.  We will discuss the issue of 
backgrounds from a new resonance at 2.46 GeV/c$^2$ when we 
describe our studies of the $D_s^{*+}\pi^0$ final state.

  From a binned maximum likelihood fit of the $\Delta M(D_s\pi^0)$ 
distribution to a Gaussian 
signal shape and second-order polynomial background function, we obtain 
a yield of $165\pm 20$ events in the peak near 350 MeV/c$^2$.  
In this fit, the mean and Gaussian width of the peak are allowed 
to float.  These parameters are determined to be 
$\langle{\Delta M}(D_s\pi^0)\rangle  =  349.4\pm 1.0\;$ MeV/c$^2$  
and $\sigma = 8.0^{+1.3}_{-1.1}\;$MeV/c$^2$, where the 
errors are due to statistics only.  The peak is somewhat 
broader than the expected mass resolution of $6.0\pm 0.3\;$MeV/c$^2$, 
determined from Monte Carlo simulations. 
The detection efficiency associated with the reconstruction of the 
full $D_{sJ}^*(2317)^+\to D_s^+\pi^0$, $D_s^+\to\phi\pi^+$, 
$\phi\to K^+K^-$ decay chain is $(9.73\pm 0.57)\,\%$ for the portion 
of the $D_{sJ}^*(2317)^+$ momentum spectrum above $3.5\;$GeV/c, 
where this efficiency does not include the $D_s$ and $\phi$ 
decay branching fractions.

Thus, we confirm the existence of a peak in the $D_s\pi^0$ 
mass spectrum that cannot be explained as reflections from 
decays of known particles.
Our measurements of the mean mass difference and width of the peak 
are consistent with the values obtained by BaBar~\cite{babar} for 
the $D_{sJ}^*(2317)^+$.   
Further discussions of the width, as well as of systematic 
errors in the measurements of the mass and width of the 
$D_{sJ}^*(2317)$ appear later in this article.

\section{Searches for \boldmath $D_{sJ}^*(2317)$ in other channels}
\label{s-searches}

  The conclusion that the $D_{sJ}^*(2317)$ is a new narrow 
resonance decaying to $D_s\pi^0$ leads to two questions:  
(1) are there other observable decay modes, and (2) might  
additional new $c\overline{s}$ resonances  
also exist in which normally suppressed decay modes such as 
$D_s^{(*)}\pi^0$ are dominant?  To answer these questions we have 
searched in the channels $D_s\gamma$, $D_s^*\gamma$, 
$D_s^*\pi^0$ and $D_s\pi^+\pi^-$.  

  If the $D_{sJ}^*(2317)$ is a $0^+$ $L=1$ $c\overline{s}$ 
meson, as has been suggested~\cite{bardeeneichtenhill}, it could 
decay via $S$- or $D$-wave to $D_s^*\gamma$,  
but would not be able to decay to $D_s\gamma$ due to parity and 
angular momentum conservation.  Consequently,  observation of one 
or both of these channels would be interesting.  
On the other hand, if neither channel is seen, this would 
not be too surprising since these are electromagnetic 
decays, and the $D_s\pi^0$ decay, while isospin violating, 
is not as severely phase-space suppressed as in the case of 
the corresponding decay of the $D_s^*$ where the 
electromagnetic decay dominates.  The BaBar data show no 
evidence for either channel, however no upper limits were 
reported on the branching ratios for these channels.  

With regard to strong decays, 
the $D_s\pi^+\pi^-$ final state is kinematically allowed and  
isospin conserving, but would be suppressed by the 
OZI rule.  This is in contrast to the $D_s\pi^0$ 
channel for which one mechanism would be decay to a $D_s$ plus a 
virtual $\eta$, with production of the $\pi^0$ via $\eta$-$\pi^0$ 
mixing~\cite{cho}.  However, angular momentum and parity conservation 
forbid the decay of a $0^+$ state to three pseudoscalars. 
Thus, observation of the $D_s\pi^+\pi^-$ channel 
would be strong evidence against the interpretation of the 
$D_{sJ}^*(2317)$ as a $0^+$ meson.  

  Finally, it is possible that the remaining $L=1$ 
$c\overline{s}$ state with $J^P = 1^+$ could also be light 
enough that decays to $D^* K$ would be kinematically forbidden.
In this case, the strong isospin-violating 
decay of this $1^+$ state to $D_s^*\pi^0$ 
could occur via $S$-wave (the electromagnetic decays to 
$D_s\gamma$ or $D_s^{*} \gamma$ 
would also be possible), and thus a narrow peak in the 
$\Delta M(D_s^*\pi^0) = M(D_s^*\pi^0) - M(D_s^*)$ 
spectrum would be a signature of such a state.

\subsection{Searches for \boldmath $D_{sJ}^*(2317)^+$ decays to
            $D_s^+\pi^+\pi^-$, $D_s^+\gamma$ and  
            $D_s^{*+}\gamma$} 

\begin{figure}
  \includegraphics*[width=4.5in]{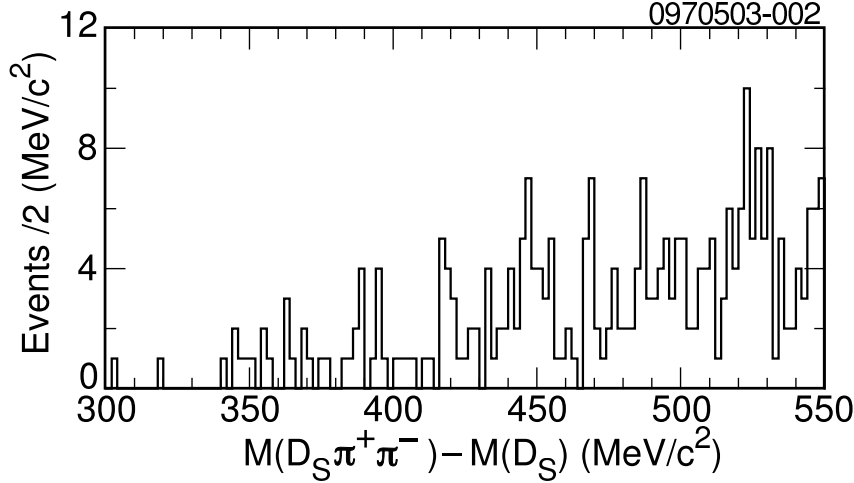}
  \caption{The mass difference 
           $\Delta M(D_s\pi\pi) = M(D_s\pi\pi) - M(D_s)$
           for $D_s^+\pi^+\pi^-$ candidates, 
           as described in the text.
           }
  \label{fig:deltampipi}
\end{figure}

  To look for these channels we select events containing 
$D_s^+\to\phi\pi^+$ candidates as in the $D_s\pi^0$ analysis.
For the $D_s\pi^+\pi^-$ channel, we combine the $D_s$ candidates 
with two oppositely charged tracks, and plot the mass difference 
$\Delta M(D_s\pi\pi) = M(D_s\pi\pi) - M(D_s)$.  
As shown in Fig.~\ref{fig:deltampipi}, no signal is evident 
in the vicinity of 350 MeV/c$^2$.  

\begin{figure}
  \includegraphics*[width=4.5in]{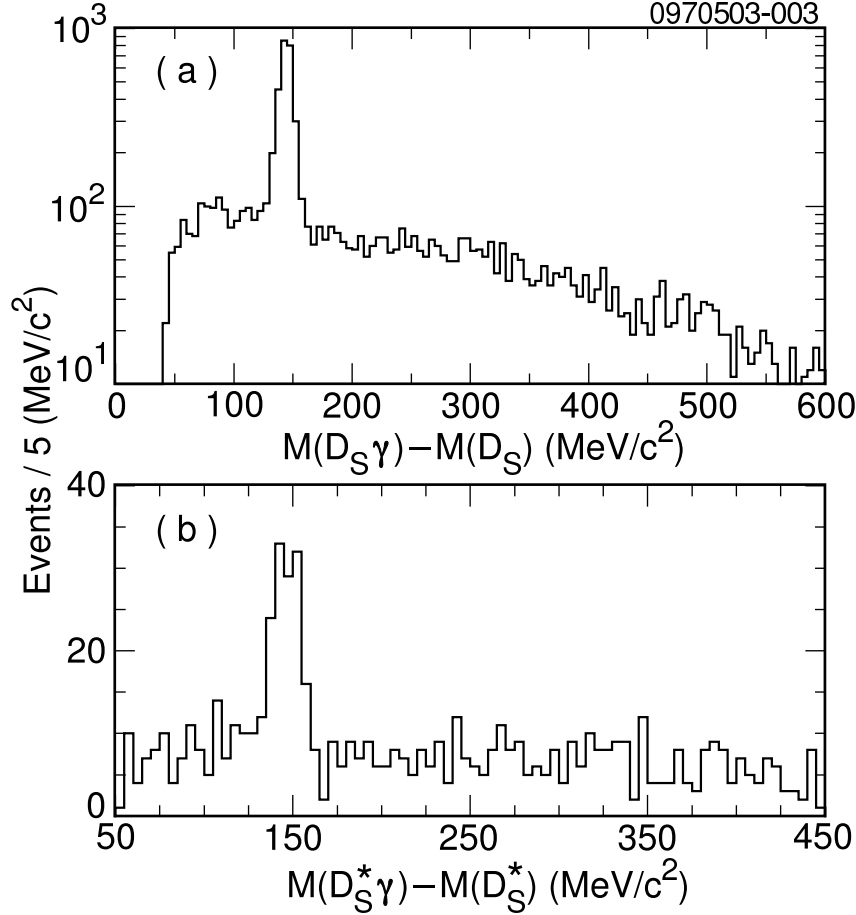}
  \caption{(a) The spectrum of the mass difference 
           $\Delta M(D_s\gamma) = M(D_s\gamma) - M(D_s)$, 
           plotted on a logarithmic scale.  The peak is due to 
           the transition $D_s^{*+} \to D_s^+ \gamma$.   
           (b) The spectrum of the mass difference 
           $\Delta M(D_s^*\gamma) = M(D_s^* \gamma) - M(D_s^*)$ 
           for $D_s^*\gamma$ candidates.
           }
  \label{fig:deltamgamma}
\end{figure}

To search for states decaying to $D_s^+ \gamma$, 
we have formed $D_s^+\gamma$ combinations by selecting photons 
of energy greater than 150 MeV. 
To select $D_s^{*+}$ candidates for use in other searches, 
we relax this to include photon candidates with energy above 50 MeV.
We ignore photons that can be paired with another photon 
such that $M(\gamma\gamma)$ is consistent with $\pi^0$ decay.
The inclusive $\Delta M(D_s\gamma) = M(D_s\gamma) - M(D_s)$ 
spectrum for this sample is plotted in Fig.\ \ref{fig:deltamgamma}(a), 
illustrating that a large $D_s^*$ sample can be obtained.  For 
decay modes with a $D_s^{*}$ in the final state, 
we select $D_s\gamma$ combinations where the mass difference 
$\Delta M(D_s\gamma)$ is reconstructed to be between 0.1308 and 0.1568 
GeV/c$^2$. 

Also visible in Fig.~\ref{fig:deltamgamma}(a),  are 
regions of the $\Delta M(D_s\gamma)$ spectrum 
where decays of the $D_{sJ}^*(2317)$ (or of 
a possible higher mass state) into $D_s\gamma$ would appear.
There is no evidence for a signal near 350 MeV/c$^2$ corresponding to 
a $M(D_s\gamma)$ in the vicinity of 2.32 GeV/c$^2$. 

The same conclusion holds for
the $D_s^*\gamma$ final state, shown in Fig.~\ref{fig:deltamgamma}(b), 
where we combine selected $D_s^*$ candidates 
with photons of energy above 150 MeV. 
The peak in the $\Delta M(D_s^*\gamma)$ spectrum in 
Fig.~\ref{fig:deltamgamma}(b) near 150 MeV/c$^2$ is due to real 
$D_s^{*+} \to D_s^+\gamma$ decays in which a random photon 
has been combined with the $D_s^+$ candidate to form the 
$D_s^*$ candidate, and the actual photon from this transition 
is combined with this system to form the $D_{sJ}^*$ candidate.  
There is no sign of any structure in this spectrum near 205 MeV/c$^2$, 
where a signal from $D_{sJ}^*(2317)$ decay would be expected.

\subsection{Search for \boldmath $D_{sJ}^*(2317)^+$ decays to
            $D_s^{*+}\pi^0$} 

\begin{figure}
  \includegraphics*[width=4.5in]{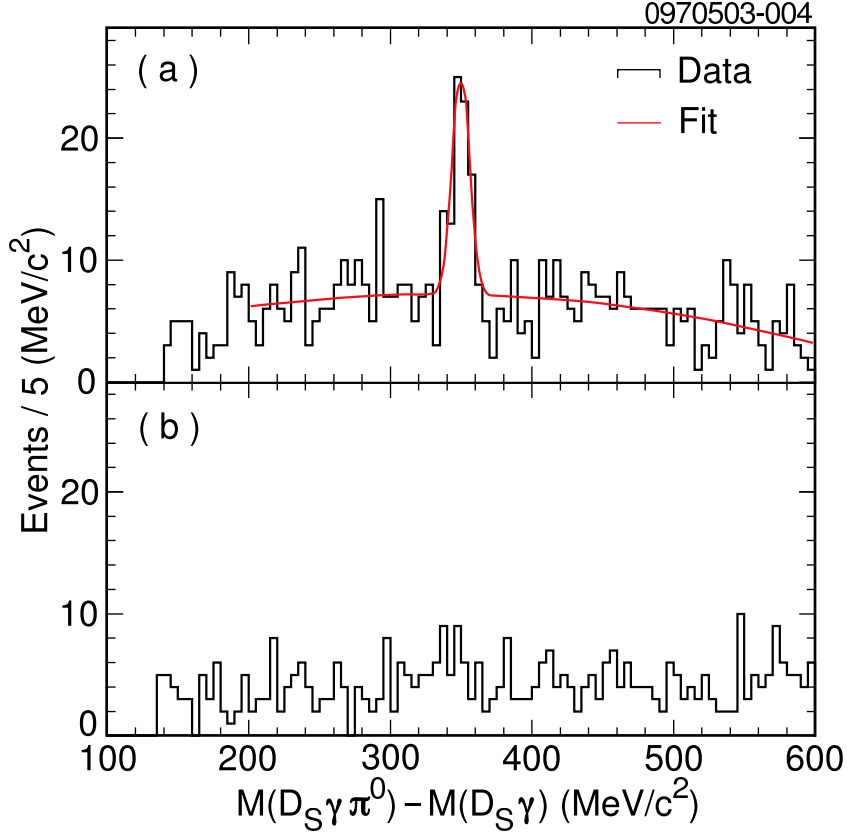}
  \caption{(a) The mass difference spectrum 
           $\Delta M(D_s^*\pi^0) = M(D_s\gamma\pi^0) - M(D_s\gamma)$
           for combinations where the $D_s\gamma$ system is consistent
           with $D_s^*$ decay, 
           as described in the text.  (b) The corresponding spectrum 
           where $D_s\gamma$ combinations are selected from the 
           $D_s^*$ side band regions, 
           defined as 
           $20.8 < |\Delta M(D_s \gamma) - 143.9\,\mbox{MeV/c}^2| < 33.8\,
           \mbox{MeV/c}^2$.
           }
  \label{fig:deltamstar}
\end{figure}

We have also searched in the $D_s^{*+}\pi^0$ channel for $D_{sJ}^*$ states. 
To maintain efficiency for this final state, 
we do not veto $D_s^{*+}$ candidates where the photon 
used in the $D_s^{*+}$ reconstruction can be combined with an 
extra photon to form a $\pi^0$ decay candidate.  
We also applied slightly less restrictive track quality 
and shower shape criteria than in the $D_s\pi^0$ analysis.  
As with the modes involving $D_s^*$ candidates described 
the preceding section, the energy of photons  
selected for reconstruction of the 
$D_s^*\to D_s\gamma$ decay is required to satisfy 
$E_\gamma > 50\;$MeV. 
The $D_s^*\pi^0$ candidates are required 
to have momenta above 3.5 GeV/c.  
Fig.~\ref{fig:deltamstar}(a) shows the mass difference plot for events 
with candidate $D_s^+\to \phi\pi^+$, $D_s^{*+}\to \gamma D_s^+$ decays 
plus di-photon combinations consistent with $\pi^0$ decay. 
If the $D_{sJ}^*(2317)^+$ 
were to decay to the $D_s^{*+}\pi^0$ final state, a peak would 
be expected at a $\Delta M(D_s^*\pi^0) \sim 205\;$MeV/c$^2$.  
Although we see no evidence for such a peak, there is a significant 
excess in a narrow region near 350 MeV/c$^2$.  
We discuss the properties of this new peak in the following section.

\section{\boldmath Observation of a new state at 
         $2.463\;\mbox{\bf GeV/c}^2$}
\label{s-observe}

From a fit to a signal Gaussian signal function plus 
a polynomial background function,  we observe 
a peak in Fig.\ \ref{fig:deltamstar}(a), comprised of 
$55 \pm 10$ $D_s^*\pi^0$ combinations, at  
$\langle{\Delta M}(D_s^*\pi^0)\rangle = 349.8 \pm 1.3\;$ MeV/c$^2$. 
The fit yields a Gaussian width of $6.1 \pm 1.0\;$MeV/c$^2$ 
for the peak, consistent with our mass 
resolution of $6.6\pm 0.5\;$MeV/c$^2$. 
The existence of this peak leads us to investigate 
the possibility of a second narrow resonance with a 
mass near 2.46 GeV/c$^2$ that decays to $D_s^{*+}\pi^0$.  
We note that a similar peak is also present in 
the $M(D_s^*\pi^0)$ spectrum observed by BaBar~\cite{babar}, 
although BaBar does not claim this as evidence for a new state.  
For ease of notation, we refer to the postulated particle
as the $D_{sJ}(2463)^+$.

\subsection{Analysis of cross feed between \boldmath $D_s^{+}\pi^0$ and 
            $D_s^{*+}\pi^0$ samples}
\label{sec:feeddown}

The kinematics of the $D_s^{+}\pi^0$ and $D_s^{*+}\pi^0$ decays 
are quite similar, and it is possible that they can reflect 
into one another.  For example, by ignoring the photon from the 
$D_s^*$ decay in $D_{sJ}(2463)^+ \to D_s^{*+}\pi^0$ decays,  
nearly all the putative signal combinations form a peak in 
the $\Delta M(D_s\pi^0)$ spectrum in the same region 
as the $D_{sJ}^*(2317)$ signal described in previous sections of this 
article.  We refer to the background entering via this scenario in the  
$D_{sJ}^*(2317)^+\to\ D_s^+\pi^0$ sample as `feed down' from the 
$D_{sJ}(2463)$ state.
The impact of neglecting the photon is that this peak in 
$\Delta M(D_s\pi^0)$ is broader than that for real 
$D_{sJ}^*(2317)$ decays.  From Monte Carlo simulations, we 
determine the width of this smeared peak to be 
$\sigma = 14.9\pm 0.4$ MeV/c$^2$.

It is also possible that a $D_s^{+}\pi^0$ candidate can be combined with 
a random photon such that the $D_s^+\gamma$ combination accidentally 
falls in the $D_s^{*+}$ signal region defined earlier.  In this case, 
$D_{sJ}^*(2317)^+\to D_s^+\pi^0$ decays would reflect or `feed up' 
into the $D_{sJ}(2463)^+\to D_s^{*+}\pi^0$ signal region. 
A Monte Carlo simulation of $D_{sJ}^*(2317)^+$ production and 
decay to $D_s^+\pi^0$ shows that this does happen, 
but only for approximately $9\%$ of the reconstructed decays.  The peak 
in the $\Delta M(D_s^*\pi^0)$ distribution generated by this feed up 
sample is also broadened relative to the expectation for real 
$D_{sJ}(2463)$ decays, analogous to the smearing of the feed down 
kinematics discussed in the preceding paragraph.  

We can extract the number of real 
$D_{sJ}^*(2317)^+\to D_s^{+}\pi^0$ decays reconstructed in our data, 
denoted as $R_0$, as well as the number of real 
$D_{sJ}(2463)^+\to D_s^{*+}\pi^0$ decays, denoted as 
$R_1$, taking into account 
that the corresponding real signal decays in one channel can enter 
the candidate sample for the other channel as described above. 
The following linear equations relate the real to observed numbers:
\begin{equation}
  N_0 = R_0 +  f_1\, R_1 
\label{eq:nzero}
\end{equation}
\begin{equation}
  N_1 = R_1 + f_0 \, R_0  \, , 
\label{eq:none}
\end{equation}
where $N_0$ and $N_1$ are the numbers of observed decays 
in the $D_s\pi^0$ and $D_s^*\pi^0$ channels respectively, and 
$R_0$ and $R_1$ are the number of real decays produced times
the efficiency to observe them in the corresponding signal 
decay channels.  The coefficients $f_0$ and $f_1$ are 
the feed up and feed down probabilities relative
to the reconstruction efficiency for the respective 
signal modes.  We note that these relations represent 
first-order approximations; higher-order corrections,  
such as that due to the scenario where 
the $D_s\pi^0$ system from a real $D_{sJ}(2463)$ decay is
combined with an unrelated photon to form a 
feed up $D_{sJ}(2463)$ candidate, are negligible 
in the present case.

The observed number of decays in the $D_s^*\pi^0$ channel 
is $N_1 =55\pm 10$, obtained from the fit to the peak 
in Fig.~\ref{fig:deltamstar}(a) described above.  
For $N_0$, it is desirable to obtain a $D_s\pi^0$ sample 
selected with criteria that most closely match those used 
to select $D_s^*\pi^0$ combinations, and that is enriched 
in $D_{sJ}^*(2317)$ decays relative to feed down from 
$D_{sJ}(2463)$ decays.  
Thus we apply the same selection criteria that were used for the 
$D_s^*\pi^0$ sample, but without selecting the photon from the 
$D_s^*\to D_s\gamma$ transition.  To measure the event yield 
in this sample, we fit the peak in the $\Delta M(D_s\pi^0)$ 
distribution to a Gaussian with its width fixed to 
the Monte Carlo expectation for $D_{sJ}^*(2317)$ decays.  
In this fit, a significant fraction of feed down combinations 
is counted as part of the combinatoric background rather than 
as signal.  
We obtain $N_0 = 190\pm 19$ candidates.  
This sample effectively constitutes the source of 
potential feed up candidates. 
The difference between this yield and the $165\pm 20$ events 
reported in Section~\ref{s-conf} is consistent with the different 
acceptances for the two sets of selection criteria.

From the Monte Carlo simulations we measure 
$f_0 = 0.091 \pm 0.007\pm 0.015$  
for the probability that a reconstructed 
$D_{sJ}^*(2317)\to D_s\pi^0$ can be combined with a 
random photon to mimic a $D_{sJ}(2463)\to D_s^*\pi^0$ decay. 
The first error is due to limited Monte Carlo 
statistics while the second is due to 
systematic uncertainties associated with: (1) the modeling of extra 
photons in the simulations, and (2) 
the fraction of such combinations that are counted by the fit 
to the $\Delta M(D_s^*\pi^0)$ distribution as contributing to 
the Gaussian signal.  As indicated above, this fraction 
counted by the fit is less 
than one due to the smearing of $\Delta M(D_s^*\pi^0)$ 
that results when an unrelated photon is added to the $D_s\pi^0$ 
system.  The agreement between the data and Monte Carlo distributions 
in Fig.~\ref{fig:manddeltam} lends confidence in the modeling of 
extra photons.  We assign a relative systematic uncertainty 
of $5\%$ based on this and on studies of combinations entering 
$D_s^*$ side bands described in the following section.  To study 
the second source of systematic uncertainty, we have carried out 
fits to the $\Delta M(D_s^*\pi^0)$ distribution in which the 
width of the Gaussian signal function was fixed to $\pm 1\sigma$
relative to the central value obtained from the nominal fit.
Based on the resulting variation in event yields, we have 
estimated a relative uncertainty on $f_0$ of $16\%$ from 
this source.

We also obtain $f_1 = 0.84 \pm 0.04 \pm 0.10$ from Monte Carlo 
simulations, where the first error is statistical and the second 
is due to systematic errors.  
This includes the probability of feed down as well 
as the photon finding efficiency.  If all $D_{sJ}(2463)\to D_s^*\pi^0$ 
decays with a reconstructed $D_s$ plus $\pi^0$ combination were to 
be counted as $D_{sJ}^*(2317)$ decays, $f_1$ would simply be one divided 
by the acceptance for finding the photon from the $D_s^*\to D_s\gamma$ 
transition.  However, because the $\Delta M(D_s\pi^0)$ distribution 
for the feed down background is broadened, a significant fraction of 
these combinations are not counted as part of the Gaussian signal, 
instead being absorbed into the polynomial background.  
The contributions to the relative systematic error on $f_1$ 
are estimated to be $5\%$ from the uncertainty on the photon-finding 
efficiency and $11\%$ from the uncertainty on the probability of 
feed down, obtained by performing alternate fits to the 
$\Delta M(D_s\pi^0)$ distribution.

Inverting Equations~\ref{eq:nzero} and \ref{eq:none}, we find that 
$R_0 = 155 \pm 23$ decays and $R_1 = 41 \pm 12$ decays, 
where the uncertainties include both statistical and systematic 
sources.  The result for $R_1$ demonstrates  
the existence of a state at 2463 MeV/c$^2$.  
The significance of the signal for this state, accounting for 
statistical and systematic errors,  
is determined to be in excess of $5\sigma$ by computing the 
probability for the combinatoric  
background plus the feed up background to fluctuate up to 
give the observed yield in the signal region in 
Fig.~\ref{fig:deltamstar}(a).

\subsection{Further evidence for the \boldmath 
            $D_{sJ}(2463)^+\to D_s^*\pi^0$ decay}
\label{sec:more2463}

We conclude from the analysis described in the preceding section
that a new state, the $D_{sJ}(2463)$, exists in addition to the 
$D_{sJ}^*(2317)$ state reported by BaBar, because feed up from the 
$D_{sJ}^*(2317)$ is only a minor background component ($\sim 25\%$) 
of the narrow peak 
observed in Fig.~\ref{fig:deltamstar}(a).  To provide further support 
for this conclusion, we have directly measured  the 
feed up background in Fig.~\ref{fig:deltamstar}(a) due to 
$D_{sJ}^*(2317)^+ \to D_s^+\pi^0$ plus random photon combinations, 
by selecting combinations 
in $D_s^*$ side band regions in the $D_s \gamma\pi^0$ 
sample. The $M(D_s\gamma\pi^0) - M(D_s\gamma)$ distribution 
for this sample, plotted in Fig.~\ref{fig:deltamstar}(b), 
shows only a small enhancement in the 
region of the $D_{sJ}(2463)$, demonstrating that the 
background from $D_{sJ}^*(2317)$ decays indeed constitutes only a  
small fraction of the entries in the $D_{sJ}(2463)$ peak.

We performed a binned likelihood fit of the spectrum in 
Fig.~\ref{fig:deltamstar}(a) to a Gaussian signal shape plus 
a second-order polynomial plus the spectrum from the $D_s^*$ 
side band region in Fig.~\ref{fig:deltamstar}(b) with its normalization 
fixed.  From this fit, we obtain $R_1 = 45.7\pm 11.6$ decays, 
consistent with the value of $R_1$ obtained from 
Equations~\ref{eq:nzero} and \ref{eq:none}.  
From the change in the likelihood of fits performed 
with and without the $D_{sJ}(2463)$ 
signal contribution, we infer that the statistical 
significance of the signal is $5.7\,\sigma$.  

Finally we note that the width of the peak in Fig.~\ref{fig:deltamstar}(a), 
$\sigma = 6.1 \pm 1.0\;$MeV/c$^2$, is consistent with the detector 
resolution.  If the origin of this peak was feed up from 
$D_{sJ}^*(2317)^+\to D_s^+\pi^0$ decays, then the effect of including 
unrelated photons to form $D_s^*\pi^0$ candidates would be 
to smear out the $\Delta M(D_s^*\pi^0)$ distribution, in the same 
way that the feed down background to the $D_{sJ}^*(2317)$ state is 
broadened as described in the preceding section.  From fits to 
Monte Carlo simulations of this feed up process, the expectation for 
the width is determined to be $\sigma = 14.9\pm 0.6$ MeV/c$^2$.  
Thus, the narrowness 
of the peak in Fig.~\ref{fig:deltamstar}(a) also rules out the possibility 
that the peak is dominantly due to feed up from $D_{sJ}^*(2317)^+$ decays.

\section{Properties of the \boldmath $D^{*}_{sJ}(2317)^+$ 
         and $D_{sJ}(2463)^+$ States}
\label{s-properties}

\subsection{\boldmath Mass and width of the $D^{*}_{sJ}(2317)^+$}

Having obtained evidence for the $D_{sJ}(2463)$ state, and 
having characterized the background that it contributes in the 
$\Delta M(D_s\pi^0)$ mass difference spectrum, we are now able to 
further address properties of the $D_{sJ}^*(2317)$ state.  
We recall that our measurement of the width of the peak 
in Fig.~\ref{fig:manddeltam} is $\sigma = 8.0^{+1.3}_{-1.1}\;$MeV/c$^2$, 
somewhat larger than our mass difference resolution,  
$\sigma = 6.0\pm 0.3\;$MeV/c$^2$.  This difference is consistent with 
predictions from Monte Carlo simulations where we include both 
$D_{sJ}(2463)$ and $D_{sJ}^*(2317)$ production, 
since roughly $18\%$ of the observed $D_s^+\pi^0$ decays 
in the $D_{sJ}^*(2317)$ signal region enter as feed down from 
the $D_{sJ}(2463)$ state, this `background' peak having an 
expected width of $\sigma = 14.9\pm 0.4\;$MeV/c$^2$.

To better determine the mass and natural width of the $D_{sJ}^*(2317)$, 
we carry out a binned likelihood fit of the peak in the 
$\Delta M(D_s\pi^0)$ spectrum in Fig.~\ref{fig:manddeltam}(b) to a 
sum of two Gaussians, one for the $D_{sJ}^*(2317)$ signal, 
and one to account for the feed down from the $D_{sJ}(2463)$.  
Allowing the means and widths of both Gaussians to float, 
we obtain $\langle{\Delta M}(D_s\pi^0)\rangle =  350.0\pm 1.2\;$MeV/c$^2$  
with $\sigma = 6.0\pm 1.2\;$MeV/c$^2$ for the $D_{sJ}^*(2317)$ 
component.  
The mean mass difference and width for the feed down component are 
$344.9\pm 6.1\;$MeV/c$^2$ and $16.5\pm 6.3\;$MeV/c$^2$, respectively.
The errors in the above values are due to statistics only;   
systematic errors are discussed below.  
Both widths are consistent with predictions from Monte Carlo 
simulations in which the two states are modeled with 
a natural width of zero.  

We have also carried out fits in which one or both of 
the widths of the Gaussians were fixed to values determined 
by the Monte Carlo.  In all cases the results were consistent 
with the results from the fit described above.  
We have also tried to obtain a purer $D_{sJ}^*(2317)$ sample 
by vetoing combinations with photons that can be combined with the 
$D_s$ candidate to form a $D_s^*$, thereby removing some 
of the feed down background from the $D_{sJ}(2463)$.  
This veto marginally improves the $D_s\pi^0$ signal 
when we fit with two Gaussians, and the mass and width 
change by only a small fraction of the statistical uncertainty.
The systematic uncertainty for $\langle{\Delta M}(D_s\pi^0)\rangle$ 
receives contributions from uncertainties in the characterization  
of the $D_{sJ}(2463)$ feed down and from uncertainties in 
the modeling of the energy resolution of the calorimeter. 
We estimate the total systematic error on the mass difference 
to be 1.0 MeV/c$^2$.  Based on these studies, we 
limit the natural width of the $D_{sJ}^*(2317)$ 
to be $\Gamma < 7\;$ MeV at the $90\%$ confidence level (C.L.).  
 
\subsection{\boldmath Mass and width of the $D_{sJ}(2463)^+$}

From the fit to the distribution resulting from the subtraction of 
Fig.~\ref{fig:deltamstar}(b) from Fig.~\ref{fig:deltamstar}(a) reported 
in Section~\ref{sec:more2463}, we obtain 
$\langle{\Delta M}(D_s^*\pi^0)\rangle = 351.2 \pm 1.7\pm 1.0\;$MeV/c$^2$    
for the mass difference between the $D_{sJ}(2463)$ and 
the $D_s^*$.  The first error is statistical and the second is 
the systematic uncertainty which is the same as that presented 
in the previous section for the $D_{sJ}^*(2317)- D_s$ mass difference.  
From our fits to data and Monte Carlo $\Delta M(D_s^*\pi^0)$ distributions, 
we also infer a $90\%$ C.L. upper limit on the 
natural width ($\Gamma$) of the $D_{sJ}(2463)^+$ state to be $7\;$MeV.

\subsection{\boldmath Production Properties}

We now give a measure of the production rates of 
$D_{sJ}^*(2317)$ and $D_{sJ}(2463)$ mesons. 
A full understanding would require the determination 
of the fragmentation functions of both particles
and their branching ratios into the final states we observe.
To minimize systematic errors, we report the relative yields
with respect to $D_s^+$ production, where all putative
charmed-antistrange systems have momenta greater than 3.5 GeV/c. 
We use all observed events for each channel, which includes direct 
production and any contributions from decays of higher mass objects. 
Then
\begin{eqnarray}
  {{\sigma\cdot {\cal {B}}\left(D_{sJ}^*(2317)\to D_s^+\pi^o\right)}
  \over{\sigma\left(D_s^+\right)}}&=&(7.9\pm 1.2\pm 0.4)\times 10^{-2},\\
  {{\sigma\cdot {\cal {B}}\left(D_{sJ}(2463)\to D_s^{*+}\pi^o\right)}
  \over{\sigma\left(D_s^+\right)}} &=& (3.5\pm 0.9\pm 0.2)\times 10^{-2},  
\end{eqnarray}

We also note that
\begin{equation}
  {{\sigma\cdot {\cal {B}}\left(D_{s}^{*+}(2112)\to D_s^+\gamma\right)}
  \over{\sigma\left(D_s^+\right)}} = 0.59\pm 0.03\pm0.01,  
\end{equation}
here and above, the first error includes the statistical and
systematic errors on the event yields while the second includes 
the systematic errors for photon detection ($2\%$), and for 
$\pi^0$ detection ($5\%$). 

\subsection{\boldmath Decays of $D_{sJ}^*(2317)$ to other final states}

With regard to the alternate $D_{sJ}^*(2317)$ decay channels 
described earlier, in which no signals were observed, 
we summarize the limits on the 
branching fractions relative to the $D_s^+\pi^0$ mode 
in Table~\ref{tab:limitszero}.  
The normalization for these limits is based on the determination 
that $(81.7\pm 5.7)\,\%$ 
of the observed yield of $165\pm 20$ entries  
in the peak of the $\Delta M(D_s\pi^0)$ spectrum 
in Fig.~\ref{fig:manddeltam}(b) are 
attributable to $D_{sJ}^*(2317)\to D_s\pi^0$ decay after 
accounting for the feed down from decays of the $D_{sJ}(2463)$ 
state to $D_s^*\pi^0$.  We have estimated the systematic error 
on this yield to be $\pm 16$ entries by varying selection criteria 
and the parameterization of signal and background shapes used in 
the fit to Fig.~\ref{fig:manddeltam}. 
\begin{table}
  \caption{\label{tab:limitszero} The 90\% C.L.\ upper limits on the ratio
  of branching fractions for $D_{sJ}^*(2317)$ to the channels 
  shown relative to the $D_s^+\pi^0$ state.  Also shown are the 
  theoretical expectations from Ref.~\cite{bardeeneichtenhill}, 
  under the assumption that the $D_{sJ}^*(2317)$ is the 
  lowest-lying $0^+$ $c\overline{s}$ meson.}
  \begin{ruledtabular}
  \begin{tabular}{lrccc}
  Final State & Yield & Efficiency & Ratio ($90\%$ C.L.) & Prediction \\
  \hline
  $D_s^+ \pi^0 $           & $135\pm 23$   & $(9.7 \pm 0.6)\,\%$ &  --- 
                           & \\
  $D_s^+ \gamma$           & $-19\pm 13$   & $(18.5 \pm 0.1)\,\%$ & $< 0.052 $ 
                           & 0 \\
  $D_s^{*+} \gamma$        & $-6.5\pm 5.2$ & $( 7.0 \pm 0.5)\,\%$ & $< 0.059 $ 
                           & 0.08 \\
  $D_s^+ \pi^+\pi^-$       & $2.0\pm 2.3$  & $(19.8 \pm 0.8)\,\%$ & $< 0.019 $ 
                           & 0 \\
  $D_s^{*+} \pi^0$         & $-1.7\pm 3.9$ & $( 3.6 \pm 0.3)\,\%$ & $< 0.11 $ 
                           & 0 \\
  \end{tabular}
  \end{ruledtabular}
\end{table}
The event yields for the various final states 
are obtained by fitting the mass difference 
distributions to Gaussians 
with each mean fixed to the result from the $D_s^+\pi^0$ channel 
and each width given by the resolution determined from 
the simulation of the corresponding decay mode.  
Uncertainties are dominated by the statistical error on the
fitted yields and limits on the relative rates
are calculated assuming a Gaussian distribution
with negative values not allowed.

\subsection{\boldmath Decays of $D_{sJ}(2463)$ to other final states}

Unlike the case of a $0^+$ state, the $D_s\pi^+\pi^-$ decay mode, 
as well as both radiative decay modes $D_s\gamma$ and $D_s^*\gamma$
are allowed for a state with $J^P = 1^+$.  From fits to the 
mass difference distributions displayed in 
Figs.~\ref{fig:deltampipi} and \ref{fig:deltamgamma} for peaks 
in the regions where a contribution from the $D_{sJ}(2463)$ would 
appear, we find no evidence of decays to any of these final states. 
We summarize the 
limits obtained on these decays, relative to $D_s^*\pi^0$, in 
Table~\ref{tab:limitsone}.
\begin{table}
  \caption{\label{tab:limitsone} The 90\% C.L.\ upper limits on the ratio
  of branching fractions for $D_{sJ}(2463)$ to the channels 
  shown relative to the $D_s^{*+}\pi^0$ state.  Also shown are the 
  theoretical expectations from Ref.~\cite{bardeeneichtenhill}, 
  under the assumption that the $D_{sJ}(2463)$ is the 
  lowest-lying $1^+$ $c\overline{s}$ meson.}
  \begin{ruledtabular}
  \begin{tabular}{lrccc}
  Final State & Yield & Efficiency & Ratio ($90\%$ C.L.) & Prediction \\
  \hline
  $D_s^{*+} \pi^0$         & $41\pm 12$    & $(6.0 \pm 0.2)\,\%$ &  --- 
                           & \\
  $D_s^+ \gamma$           & $ 40\pm 17$   & $(19.8 \pm 0.4)\,\%$ & $< 0.49 $ 
                           & 0.24 \\
  $D_s^{*+} \gamma$        & $-5.1\pm 7.7$ & $( 9.1 \pm 0.3)\,\%$ & $< 0.16 $ 
                           & 0.22 \\
  $D_s^+ \pi^+\pi^-$       & $2.5\pm 5.4$  & $(19.5 \pm 1.5)\,\%$ & $< 0.08 $ 
                           & 0.20 \\
  $D_{sJ}^*(2317)^+\gamma$ & $3.6\pm 3.0$  & $(2.0 \pm 0.1)\,\%$  & $< 0.58 $ 
                           & 0.13 \\
  \end{tabular}
  \end{ruledtabular}
\end{table}

Despite a high relative efficiency, the limit on the decay 
$D_{sJ}(2463)^+\to D_s^+ \gamma$ is less stringent than those on the 
decays to $D_s^{*^+}\gamma$ and $D_s^+\pi^+\pi^-$.  This is due to an 
excess of combinations in the signal region.  From fits performed with 
and without the signal Gaussian, we determine that the statistical 
significance of this excess is 2.4 standard deviations.

\begin{figure}
  \includegraphics*[width=4.5in]{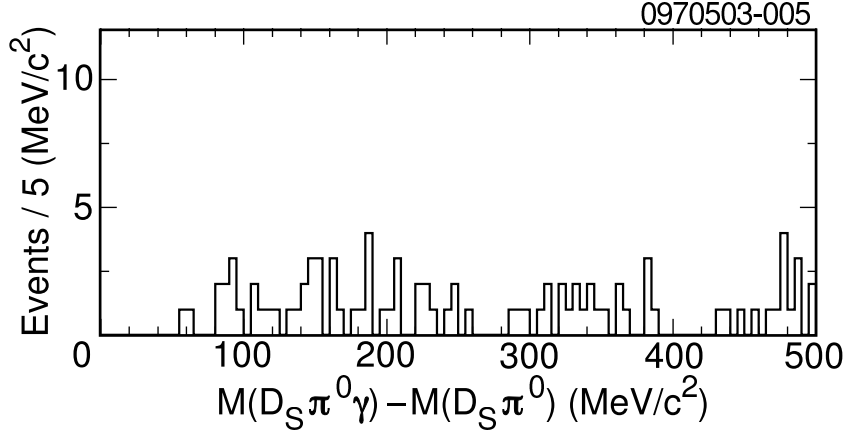}
  \caption{The mass difference spectrum 
           $M(D_s\pi^0\gamma) - M(D_s\pi^0)$
           for candidates for the decay 
           $D_{sJ}(2463)\to D_{sJ}^*(2317)\gamma$, after 
           application of 
           the selection criteria described in the text.  
           }
  \label{fig:deltam2317}
\end{figure}

If the $D_{sJ}(2463)^+$ is a $1^+$ state, then it is also possible for it 
to undergo a P-wave radiative decay to 
$D_{sJ}^*(2317)^+\gamma$~\cite{noteonxfeed}. 
We have looked for this transition in our $D_s\gamma\pi^0$ sample.
To reduce backgrounds from $D_{sJ}(2463)^+\to D_s^{*+}\gamma$, we 
required that the $D_s\pi^0$ system be consistent with the decay 
of the $D_{sJ}^*(2317)$, namely that 
$|\Delta M(D_s\pi^0) - 350.0\,\mbox{\rm MeV/c}^2| < 13.4\,$MeV/c$^2$ 
($\sim 2\sigma$ based on Monte Carlo simulations).
We also required that the $D_s\gamma$ system be inconsistent with 
$D_s^*$ decay at the $1\sigma$ level (the corresponding 
$\Delta M(D_s\gamma)$ 
must deviate from the expected value for this decay by 
more than $4.4\;$MeV/c$^2$), 
and that the momentum of the $\pi^0$ be inconsistent with the 
$D_{sJ}(2463)\to D_s^*\pi^0$ transistion, also at the $1\sigma$ level.
The $M(D_s\pi^0\gamma) - M(D_s\pi^0)$ distribution, plotted in 
Fig.~\ref{fig:deltam2317}, provides no evidence for a signal 
in the vicinity of 150 MeV/c$^2$.
Due to the tightness of these cuts, the efficiency 
for detecting this decay is roughly a factor of three smaller than 
for the $D_s^*\pi^0$ decay mode.  
The $90\%$ C.L.\ 
upper limit for this channel is reported in the bottom row of 
Table~\ref{tab:limitsone}.

\section{Conclusions and Discussion}
\label{s-conclude}

  In summary, data from the CLEO~II detector have provided confirming 
evidence for the existence of a new narrow resonance decaying to 
$D_s^+\pi^0$, with a mass near 2.32 GeV/c$^2$.  This state is consistent 
with being the $0^+$ member of the lowest-lying $P$-wave $c\overline{s}$ 
multiplet.  As summarized in Table~\ref{tab:limitszero},   
we have set upper limits on other decay modes of this state.  
We have measured the mass splitting 
of this state with respect to the $D_s$ meson to be 
$ 350.0\pm 1.2\;\mbox{\rm [stat.]}\pm 1.0\;\mbox{\rm [syst.]}\;$MeV/c$^2$, 
and we find its natural width to be $\Gamma < 7\;$MeV at $90\%$ C.L.

We have observed and established the existence of a new 
narrow state with a mass near 
2.46 GeV/c$^2$ in its decay to $D_s^{*+}\pi^0$, 
which we have denoted $D_{sJ}(2463)$.  We have demonstrated that
the signal for this decay cannot be interpreted as a reflection 
from the $D_{sJ}^*(2317)^+\to D_s^+\pi^0$ decay.  
The measured properties of this state are consistent with its 
interpretation as 
the $1^+$ partner of the $0^+$ state in the spin multiplet with 
light quark angular momentum of $j=1/2$.  
We have measured the mass splitting of this 
state with respect to the $D_s^*$ meson to be 
$ 351.2\pm 1.7\;\mbox{\rm [stat.]}\pm 1.0\;\mbox{\rm [syst.]}\;$MeV/c$^2$.
The natural width of this state is found to be $\Gamma < 7\;$MeV 
at $90\%$ C.L.  Since the $D_{sJ}(2463)$ mass lies above the 
kinematic threshold for decay to $DK$ (but not for $D^*K$), the narrow 
width suggests this decay does not occur. 
Since angular momentum and parity conservation laws forbid a $1^+$ state 
from decaying to two pseudoscalars, this provides additional 
evidence for the compatibility of the $D_{sJ}(2463)$ with the 
$J^P = 1^+$ hypothesis.  

In the model of Bardeen, Eichten and Hill~\cite{bardeeneichtenhill}, 
a $J^P = 1^+$ state is predicted with the same mass splitting $\Delta M$ 
with respect to the $1^-$ state as that between the $0^+$ and $0^-$ states.  
Taking the difference between the two mean mass differences reported 
above, we obtain $\delta(\Delta M) = (351.2\pm 1.7) - (350.0\pm 1.2) = 
1.2 \pm 2.1\;$MeV/c$^2$ for the difference between the
$1^+ - 1^-$ and $0^+ - 0^-$ mass splittings, where the dominant 
uncertainty is due to statistics.  Thus our observations are 
consistent with these predictions.  


We gratefully acknowledge the effort of the CESR staff 
in providing us with excellent luminosity and running conditions.
We thank W.\ Bardeen, E.\ Eichten, S.\ Godfrey, C.\ Hill  
and J.\ Rosner for useful discussions.
M. Selen thanks the Research Corporation, 
and A.H. Mahmood thanks the Texas Advanced Research Program.
This work was supported by the 
National Science Foundation, 
and the
U.S. Department of Energy.

\clearpage


\begin{thebibliography}{99}
\bibitem{noteonxfeed}  The calculation of cross feed backgrounds in 
                       Section~\ref{sec:feeddown} did not account for 
                       the possibility of 
                       $D_{sJ}(2463)\to D_{sJ}(2317) \gamma$ decays.  
                       However, the impact of ignoring this channel is 
                       insignificant based on our upper limit.
\end{thebibliography}
\end{document}